\documentclass[aps, prl, reprint, floatfix, amsmath, notitlepage, tightenlines, superscriptaddress]{revtex4-1}

\usepackage{graphicx}
\usepackage{physics}
\usepackage{float}
\usepackage[colorlinks,allcolors=blue]{hyperref}
\usepackage{mathtools}
\usepackage{amsmath}
\usepackage{amssymb}
\usepackage{bm}

\begin{document}
\title{Dynamical formation and interaction-induced stabilization of dark condensates of dipolar excitons}

\author{Yotam Mazuz-Harpaz}
\affiliation{Racah Institute of Physics, Hebrew University of Jerusalem, Jerusalem 91904, Israel}
\author{Maxim Khodas}
\affiliation{Racah Institute of Physics, Hebrew University of Jerusalem, Jerusalem 91904, Israel}
\author{Ronen Rapaport} \email{ronenr@phys.huji.ac.il}
\affiliation{Racah Institute of Physics, Hebrew University of Jerusalem, Jerusalem 91904, Israel}
\affiliation{Applied Physics Department, Hebrew University of Jerusalem, Jerusalem 91904, Israel}

\date{\today}

\begin{abstract}
The formation of a dense Bose-Einstein condensate in dark spin states of two-dimensional dipolar excitons is shown to be driven by a dynamical transition to the long-lived dark states. The condensate is stabilized by strong dipole-dipole interactions up to densities high enough for a dark quantum liquid to form. The persistence of dark condensation was observed in recent experiments. A model describing the non-equilibrium dynamics of externally driven coupled dark and bright condensates reproduces the step-like dependence of the exciton density on the pump power or on temperature. This unique condensate dynamics demonstrates the possibility of observing new unexpected collective phenomena in coupled condensed Bose systems, where the particle number is not conserved.
\end{abstract}

\maketitle

\textbf{Introduction:}  Quantum fluids with long range, anisotropic interactions display rich variety of emergent collective phenomena. A prominent example is the dipole-dipole interaction, which has recently been addressed by growing communities, both in atomic \cite{baranov_condensed_2012, lahaye_physics_2009, kadau_observing_2016, ferrier-barbut_observation_2016, chomaz_quantum-fluctuation-driven_2016} and condensed matter physics, the latter focusing on dipolar quantum fluids of two-dimensional indirect excitons (IXs) in semiconductor double quantum wells (DQWs) \cite{high_condensation_2012,high_spontaneous_2012}. IXs have unique properties which offer opportunities to explore new physical phenomena that are currently inaccessible with atomic gases: (a) their large electrical dipole gives access to strong interaction regimes leading to observable particle correlations, short range order, and dipolar liquidity \cite{laikhtman_exciton_2009, laikhtman_correlations_2009, shilo_particle_2013, stern_exciton_2014, cohen_dark_2016}, (b) unlike atoms, the number of IXs is not conserved, and is rather determined by the interplay of drive and relaxation in the system, and (c) the phases of IXs are also intimately related to their non-zero internal spin states \cite{shilo_particle_2013, cohen_dark_2016, mazuz-harpaz_condensation_2017, stern_exciton_2014, misra_experimental_2018, anankine_quantized_2017,  high_spin_2013,beian_spectroscopic_2017}. In the optically active ("bright") states of IXs the spin projection along the dipole is $S=\pm 1$ while the states with $S=\pm 2$ are optically inactive ("dark"). For IXs in typical GaAs DQWs, the dark spin states are lower in energy than the bright states by an energy splitting $\varepsilon_{bd}$, which is of the order of a few to few tens of $\mu$eVs, and thus an IX Bose-Einstein condensate (BEC) phase was predicted to spontaneously form in the dark spin states if IXs are considered non-interacting \cite{combescot_bose-einstein_2007}. 

However, including short-range exchange interactions of electrons or holes between colliding IXs, should cause a brightening instability of the dark BEC as the density exceeds the critical value $n_{c2}$ \cite{combescot_``gray_2012}. For condensate densities $n>n_{c2}$, the BEC becomes a coherent mixture of dark and bright components, whose respective weights tend to become equal as $n\gg n_{c2}$, where it essentially becomes a bright BEC. Notably, a purely dark BEC of unpolarized excitons was evaluated to be stable only at very low densities, $n<n_{c2}\sim 10^{9}$cm$^{-2}$, which should make its observation very challenging experimentally. Nevertheless, two independent experiments on IXs with large electric dipoles \footnote{The two independent experiments which reported the observation of that robust dark liquid were performed on IXs with different and relatively large dipole lengths: $16$nm \cite{cohen_dark_2016} and $18$nm \cite{stern_exciton_2014,misra_experimental_2018}} gave strong evidence for a spontaneous formation of a collective dark IX state at densities far exceeding the aforementioned prediction of $n_{c2}$ \cite{shilo_particle_2013, stern_exciton_2014, cohen_dark_2016, mazuz-harpaz_condensation_2017, misra_experimental_2018}. The dark phase persisted up to densities corresponding to a correlated liquid state with evidence of short range order and reduced density fluctuations. Furthermore, Refs.~\citenum{cohen_dark_2016,mazuz-harpaz_condensation_2017} reported a peculiar step-like density increase as a function of both optical pump power and temperature, with the added particles being predominantly dark. 

Here we show that the long range dipolar repulsion stabilizes the dark BEC phase by suppressing the bright-dark exchange coupling up to densities high enough to support a dark quantum liquid with short range order. Remarkably, in the quantum liquid $n_{c2}$ is weakly dependent on the dipole size and on $\varepsilon_{bd}$, in agreement with the recent experiments. We develop a rate equation model describing the non-equilibrium dynamics of a driven condensate with a variable number of particles, which reproduces the sharp increase in the total number of particles at the onset of a transition to a dark BEC, as well as the sharp slowdown of the particle accumulation above a second critical density, where the bright-dark mixing occurs, as was observed in experiments.\vspace*{0.1cm}

\textbf{Correlation-induced stability of dark dipolar BEC:} Assuming local thermal equilibrium, the number of particles in the IX BEC is \cite{pitaevskii_boseeinstein_2003}
\begin{equation} \label{eq:BEC_dist}
N = \bar{N}-B(T,\varepsilon_{bd}),
\end{equation}
where $\bar{N}$ is the total number of particles, and the thermal cloud occupation number, $B(T,\varepsilon_{bd})$, is determined by the IX temperature $T$ and by $\varepsilon_{bd}$.
\footnote{In a two-dimensional parabolic trap and for a non-interacting system, $B(T,\varepsilon_{bd})= A(T/\varepsilon_{bd} )^2$, where $A$ is a constant, and we note that in the case of interactions, the transition becomes of a BKT type into a superfluid state, with a different temperature dependence. In this work however, we do not discuss the details of the 
phase transition into the condensed phase, but rather assume such a phase has been reached.}. Eq. \ref{eq:BEC_dist} is valid for $\bar{N}>\bar{N}_{c1}(T)= B(T,\varepsilon_{bd} )$ \cite{pitaevskii_boseeinstein_2003}. As the occupation further increases the Josephson coupling between the bright and dark BECs grows and causes a second, brightening transition, at $N=N_{c2}$ \cite{combescot_``gray_2012}. For $N>N_{c2}$, the occupation numbers $N_D$, $N_B$ of the dark and bright BEC components, respectively, are given by
\begin{equation} \label{eq:mixed_dist}
	N_{D,B}=(N\pm N_{c2})/2
\end{equation} 
and
\begin{equation}\label{eq:Nc2}
	N_{c2} \simeq \varepsilon_{bd} / \xi,
\end{equation}
where $\xi$ is the first order energy correction due to the exchange processes between all constituents in the binary collision of excitons. The considered exchange processes transform a pair of dark excitons into a pair of bright ones and vice versa. Assuming the holes are much heavier than electrons and using the translational invariance,
\begin{align} \label{eq:xi_general}
\begin{split}
\xi = & \int d^2\bm{r}_{e} d^2\bm{r}_{e'} d^2\bm{r} \; \psi(\bm{r}_{e},\bm{r}/2) \psi(\bm{r}_{e'},-\bm{r}/2)   \\
	  & \times \psi(\bm{r}_{e'},\bm{r}/2) \psi(\bm{r}_{e},-\bm{r}/2) \, \left| \Phi(\bm{r}) \right|^2 
        V_{XX}
\end{split}
\end{align}
where $\psi(\bm{r}_e,\bm{r}_h)=\frac{1}{\sqrt{\pi a_X^2}}\exp \left(-\frac{\left|\bm{r}_h-\bm{r}_e \right|^2}{2a_X^2} \right)$ is the wave-function of the relative motion of the $e$ and the $h$ of a given exciton, $\bm{r}$ is the relative position vector between the two holes, $\Phi(\bm{r})$ is the wave-function of the relative motion of the centers of masses of the two excitons. For an IX with its $e$ and the $h$ separated perpendicular to the plane of motion with an $e\text{-}h$ separation $d$. The Coulomb interaction between the two IXs is  
\begin{align}\label{eq:Vxx}
    V_{XX}
    = \frac{e^2}{\kappa} &
     \left[   \frac{1}{\sqrt{\left|\bm{r}_{e}-\bm{r}_{e'} \right|^2} } + \frac{1}{r} 
     - \frac{1}{\sqrt{d^2+\left|\bm{r}_{e}+\bm{r}/2 \right|^2 }}  \right. \nonumber \\
    & \left. - \frac{1}{\sqrt{d^2+\left|\bm{r}_{e'}-\bm{r}/2 \right|^2 }} \right],
\end{align}
where $\kappa$ is the dielectric constant. 
The main contribution to the exchange integral, Eq.~\ref{eq:xi_general}, comes from inter-exciton distances $r \lesssim a_X$, where the overlap between the wave-functions of the two excitons is significant. Unpolarized excitons with $d=0$ are neutral and are almost uncorrelated, i.e., $\Phi$ is practically featureless, $\left|\Phi(\bm{r})\right|^2\sim 1/L^2$, where $L$ is the size of the system. This results in a non-negligible overlap probability between two excitons, of the order of $(a_X/L)^2$, resulting in a large contribution to $\xi$ and leading to the aforementioned low, and hard to reach value of $n_{c2}$ \cite{combescot_``gray_2012}. 

The above picture should be revised, however, owing to the long range dipolar repulsion, which causes adjacent IXs to avoid getting close to each other. These experimentally observable correlations \cite{shilo_particle_2013} are reflected in a non-trivial spatial dependence of $\Phi$ and to the formation of a significant depletion region around each IX \cite{zimmermann_excitonexciton_2007, laikhtman_exciton_2009, laikhtman_correlations_2009}. 
For large enough dipoles, the depletion range may easily exceed $a_X$. In this case $\xi$ is strongly suppressed, resulting in a significantly larger $n_{c2}$. 

\begin{figure}
\includegraphics[width=0.48\textwidth]{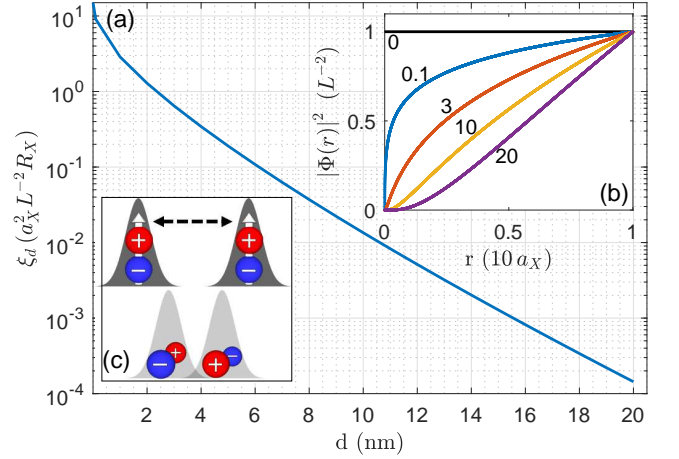}
\caption{\textbf{Interaction-induced particle correlations and suppressed exchange interaction between excitons.} (a)  The calculated exchange integral between the constituents of two IXs, $\xi_d$, as a function of the dipole length $d$, for IXs in typical GaAs DQWs (here we use $a_X=10\,nm$, $b=3\,nm$) with $k=(10a_X)^{-1}=\mu m^{-1}$. $R_X=e^2/(2\kappa a_x)$. (b) The squared wave-function of the relative center-of-mass position, $|\Phi_d(r)|^2$, of Eq.~\ref{eq:Phi_c}, for the same values of $a_X$ and $b$ as in (a). The different colors are for different dipole lengths, given in nm. (c) An illustration of the effect of long range dipolar repulsion on the suppression of $\xi$.} 
\label{fig:xi_and_phi}
\end{figure}

\begin{figure}
\includegraphics[width=0.48\textwidth]{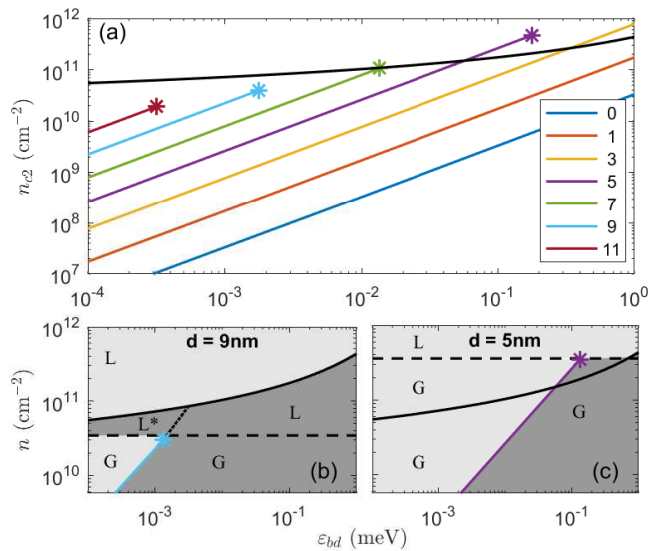}
\caption{\textbf{Second critical density of the interacting condensate and phase diagrams.} (a) The second critical condensate density $n_{c2}$ of IXs in typical GaAs DQWs (here too $a_X=10nm$, $b=3nm$) as a function of $\varepsilon_{bd}$, for different dipole lengths, from the two-body scattering model in the dilute gas limit. Each line is terminated at the density $n_{liq}$, marked by a star. The black line is the prediction of Eq.~\ref{eq:nc2_fixed} for the liquid limit with short range order. In this limit $n_{c2}$ is essentially independent of the dipole length. (b) and (c) show the resulting phase diagram in the $(n,\varepsilon_{bd})$ plane, with $n=N/L^2$. The dashed line marks the approximate gas-liquid transition density $n_{liq}$. The four different condensate regimes are a dilute gas dark-condensate (dark gray G region), a liquid dark-condensate (dark-gray L region), a dilute gas mixed-condensate (light gray G region), and a liquid mixed-condensate (light gray L liquid). The area denoted by L* is the predicted re-entrance of the dark phase due to the development of short range order.} \label{fig:nc2}
\end{figure}

To show this, we use the solution for $\Phi_d(r)$ from the quantum scattering problem of two dipolar IXs, developed in Ref.~\citenum{laikhtman_exciton_2009},
\begin{equation} \label{eq:Phi_c}
	\Phi_d (r) = \frac{1}{L}
    \begin{cases}
    	\frac{K_0 (2d/\sqrt{br})}{K_0 (2d/\sqrt{bk^{-1}})} & r<k^{-1} \\
    	1 & r \ge k^{-1},
    \end{cases}
\end{equation}
where $K_0$ is the modified Bessel function of the 2nd kind, $k$ is the scattering wave number, $b=\hbar^2 \kappa / Me^2$, and $M$ is the IX mass \footnote{Defined this way, $\Phi_d$ is approximately normalized over the entire area of the system $L^2$, neglecting the effect of the scattering on the wave-function at distances exceeding $k^{-1}$}. $|\Phi_d(r)|^2$ is plotted in Fig.~\ref{fig:xi_and_phi}(b) for different values of $d$. As expected, the probability of finding two excitons at distances $\sim a_X$ apart is strongly suppressed even for IXs with small $d$, and it decreases dramatically with increasing $d$. Using Eq. \ref{eq:Phi_c}, we numerically evaluate the exchange integral $\xi_d$ given by Eq.~\ref{eq:xi_general} for IXs in GaAs DQWs for various realistic values of $d$, as plotted in Fig.~\ref{fig:xi_and_phi}(a) \footnote{For IXs in a condensate the typical wavelength is of the order of $2\pi/k\sim L\gg a_X$. In addition, the dependence of $\Phi_d$ on $k$ is logarithmic and therefore weak. Therefore, for $2\pi/k\gg a_X$ \cite{laikhtman_exciton_2009}, we can safely choose $k^{-1}=10 a_X$ to calculate $\xi_d$ as a function of $d$. See supplementary note S1.}. $\xi_d$ drops exponentially with increasing $d$, reflecting the strong suppression of the short range exchange interactions. The critical  density $n_{c2}$ can now be estimated from Eq.~\ref{eq:Nc2} and as $\varepsilon_{bd}$ is a single-particle property that depends only on material parameters \cite{pikus_exchange_1971,chen_exchange_1988}, Fig.~\ref{fig:nc2}(a) presents $n_{c2}$ as a function of $\varepsilon_{bd}$. $n_{c2}$ increases dramatically with $d$: for a typical experimental value of $d=12$nm, $n_{c2}$ is larger by a factor of more than $10^4$ compared to the unpolarized case, 
$d=0$.

The above model applies only in the dilute limit, where multi-particle interactions can be neglected and only two-body scattering is considered. As can be seen from Fig.~\ref{fig:xi_and_phi}(b), the depletion region around each IX increases with $d$. When the typical depletion region reaches the average inter-particle distance, multi-particle correlations are expected, and the system is predicted to be in the liquid regime with short-range order \cite{laikhtman_exciton_2009}, as was indeed recently observed in experiments \cite{stern_exciton_2014, cohen_dark_2016}. It was shown that there exists a density regime where the liquid is dark \cite{cohen_dark_2016,misra_experimental_2018}. The transition to the liquid regime was estimated in Ref.~\citenum{laikhtman_exciton_2009} to occur at densities larger than $n_{liq}\sim b^2/(4d^4)$ and accordingly, once  $n_{c2}\sim n_{liq}$, the dilute-limit approximation breaks down. These densities are marked by the stars in Fig \ref{fig:nc2}(a), for the different values of $d$. To estimate $n_{c2}$ in the dipolar liquid we employ the qualitative description of short range order formulated in  Ref.~\citenum{laikhtman_exciton_2009}. On the qualitative description of a dipolar liquid with short range order, where each IX is considered as localized in a nearly parabolic potential defined by its neighbors. Denote the positions of the nearest neighbors of a given IX by $\bm{r}_i$, $i=1,\ldots,m$. The short range order implies $r_i \propto 1/\sqrt{n}$. In this approximation, $\Phi_d$ takes the form
\begin{equation} \label{eq:Phi_fixed}
	\Phi_d(\bm{r}) = L^{-1} \sum_i^m\delta \left( \bm{r}-\bm{r_i} \right).
\end{equation} 
Assuming Eq.~\ref{eq:Nc2} holds in a liquid, 
\begin{equation} \label{eq:nc2_fixed}
	n_{c2} L^2 \simeq \varepsilon_{bd}/ \left[m\xi_d(n_{c2})\right].
\end{equation}
in which case $\xi_d$ in Eq.~\ref{eq:xi_general} becomes a function of $n$ \footnote{We verified that the contribution from the exciton's neighbors other than nearest is negligible, since the exchange integral decays exponentially with distance, $r$ on the length scale of the IX radius.}. $\xi_d$ strongly depends on $n$ such that even large variation in the numerical factors of the r.h.s of Eq.~\ref{eq:nc2_fixed} are compensated by exponentially small variations in $n$. Therefore, $n_{c2}$ is robust against variations of $\varepsilon_{bd}$, coordination number $m$, and most importantly, of the dipole length $d$. This robustness is clearly demonstrated in the numerical solutions of Eq.~\ref{eq:nc2_fixed}, presented as the solid black curve in Fig.~\ref{fig:nc2}(a), for $m=1$. The solutions show a variation of less than a factor of 5 in $n_{c2}$, over four orders of magnitude variation in $\varepsilon_{bd}$ and no significant dependence on $d$. This is a remarkable consequence of the short range order of a dipolar liquid. Unlike in a dilute gas, in a liquid-like dark condensate, $n_{c2}$ is insensitive to any local fluctuations that do not affect the in-plane extent of the single IX wave-function. \footnote{We note that some dependence on $d$ should appear if the harmonic approximation for $\Phi_d$ from Ref.~\citenum{laikhtman_exciton_2009} instead of the Delta function of Eq.~\ref{eq:Phi_fixed}.}

Based on the above analysis, it is now possible to construct a phase diagram for the Bose-condensed dipolar IX system, i.e., for the case of $\bar{N}>\bar{N}_{c1}$. This is presented in  Fig.~\ref{fig:nc2}(b,c) for the cases of $d=9$nm and $d=5$nm, respectively. The four different condensate regimes are: a dilute gas dark condensate (dark G region), a liquid-like dark condensate (dark L region), a dilute gas mixed condensate (light G region), and a liquid-like mixed condensate (light L region). The dashed black line marks the approximate density of the gas-liquid transition, $n_{liq}$. As can be seen, dark phases are expected over a wide range of dark-bright splittings for IX systems with relatively large dipoles. 
For large enough $d$ and small $\varepsilon_{bd}$, a re-entrance transition from a mixed dilute gas condensate to a dark liquid with \textit{increasing} $n$ is predicted, which is a result of the onset of the short range order leading to suppressed exchange bright-dark mixing. This will be apparent as a darkening of an already coherent bright BEC as the density increases.
\vspace*{0.1cm}

\begin{figure*}
\includegraphics[width=\textwidth]{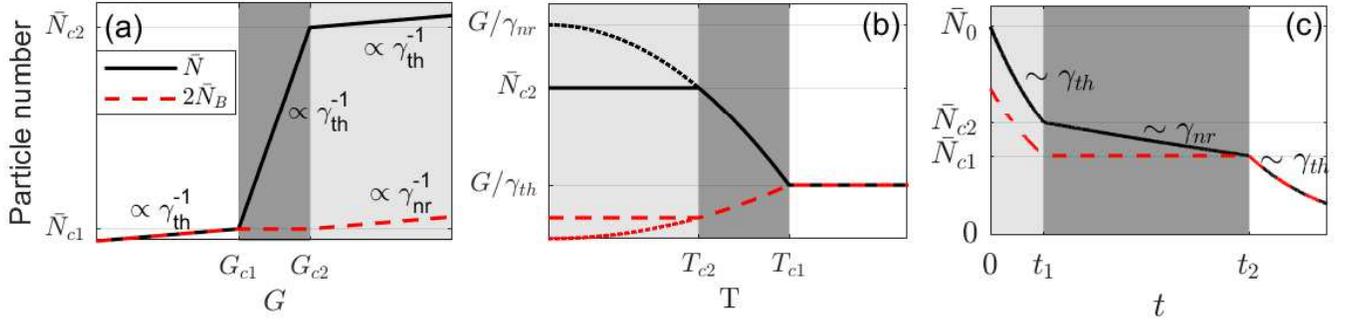}
\caption{\textbf{Steady-state and decay dynamics:} Illustrations of the total number of particles $\bar{N}$ (black line) and of $2 \bar{N}_B=2N_B+N_{th}$, which is twice the total number of IXs in bright spin states (both thermal and condensed), proportional to the expected photoluminescence from the system. (a) At steady state, as a function of the generation rate, according to Eq.~\ref{eq:N_steady_state_solution}, for the case that $T<T_{c1}$. The different linear slopes are also marked. (b) At steady state, as a function of the temperature, for the case that $G>\gamma_{nr}\bar{N}_{c2}$. The dotted lines are for the case of $G<\gamma_{nr}\bar{N}_{c2}$, where the condensate never reaches the occupancy condition to become mixed. (c) As a function of time during the decay of an initial population $\bar{N}_0$, according to supplementary Eq. S5. The background colors mark the different regimes: thermal gas (white), dark condensate (dark gray), and mixed condensate (light gray). Note that for ease of presentation the figures are not to scale.} \label{fig:dynamics}
\end{figure*}

\textbf{Dynamical high density dark condensation:} With the above picture of collective ground states of the IX fluid, we now describe the dynamics of a realistic dipolar IX system, where IXs are generated by an external pump at a rate $G$ and decay either radiatively or non-radiatively with rates $\gamma_r$ and $\gamma_{nr}$, respectively. In most experiments done on IXs in direct bandgap systems the radiative decay is much faster than non-radiative processes, i.e., $\gamma_r\gg\gamma_{nr}$. This inequality has striking consequences upon condensation. Since the thermalization of IXs between themselves and with the lattice is the fastest timescale, we assume a quasi-equilibrium dynamics in which the condensate and the thermal cloud are always in thermal equilibrium, while the number of IXs in the system obeys the rate equation 
\begin{equation} \label{eq:general_rate}
	\frac{d}{dt}\bar{N}=G - \gamma_{th} N_{th} - \gamma N,
\end{equation}
where $N_{th}$ is the occupation number of the thermal cloud and $\gamma_{th}$ and $\gamma$ are the decay rates of IXs in the thermal cloud and in the condensate, respectively. Since the temperature is typically much larger than $\varepsilon_{bd}$, the occupations of dark and bright states in the thermal cloud are nearly equal leading to an overall decay rate of the thermal cloud: $\gamma_{th} = \gamma_{r}/2 + \gamma_{nr}\simeq \gamma_{r}/2$. In contrast, $\gamma$ depends on the distribution between spin states in the condensate  and therefore also on $N$. For a purely dark condensate, the decay rate is just the non-radiative rate, $\gamma=\gamma_{nr}$, while for a mixed condensate the decay rate is determined by the relative weights of the dark and bright components of the condensate, Eq.~\ref{eq:mixed_dist}, leading to $\gamma=\left[N_D\gamma_{nr}+(\gamma_{nr}+\gamma_r)N_B\right]/N$.

In steady state $d\bar{N}/dt=0$ and thus, using the obtained decay rates and Eq.~\ref{eq:BEC_dist}, one can solve Eq.~\ref{eq:general_rate} and find the total IX number for each of the three phases:
\begin{equation} \label{eq:N_steady_state_solution}
\bar{N}=
\begin{cases}
	G/\gamma_{th} & \bar{N}<\bar{N}_{c1} \\
     G/\gamma_{nr} - \frac{1}{2}\frac{\gamma_r}{\gamma_{nr}}\bar{N}_{c1}(T) & \bar{N}_{c1}<\bar{N}<\bar{N}_{c2} \\
     G/\gamma_{th} +\frac{1}{2}\frac{\gamma_r}{\gamma_{th}}N_{c2} & \bar{N}_{c2}<\bar{N}
\end{cases}
\end{equation}
where $\bar{N}_{c2}\equiv \bar{N}_{c1}+N_{c2}$ is the total number if particles at the transition from dark to mixed condensate and the corresponding occupation number of the condensate is everywhere given by Eq. \ref{eq:BEC_dist}. 
This solution predicts a dramatic dynamical effect: starting from a thermal cloud of IXs that decay mostly radiatively with a rate $\gamma_{th}$, as the excitation power is increased or the temperature is decreased to the point where $\bar{N}=\bar{N}_{c1}(T)$, a dark IX condensate forms and its decay is now governed by the much smaller $\gamma_{nr}$. This will result in a sharp dynamical increase of the IX cloud density, driven by major accumulation of IXs in the dark condensate. This accumulation only stops when $N=N_{c2}$ and the condensate becomes mixed, resulting in the reactivation of the fast radiative channel. This unique dynamics result in a distinct step-like density profile as a function of either $G$ or $T$, as is illustrated in Fig.~\ref{fig:dynamics}(a,b). The critical particle generation rates corresponding to the two critical densities are
\begin{subequations} \label{eq:Gc}
\begin{align}
	G_{c1} &= \gamma_{th} \bar{N}_{c1}(T), \\
    G_{c2} &= \gamma_{th} \bar{N}_{c1}(T) + \gamma_{nr}N_{c2},
\end{align}
\end{subequations}
and similarly, the corresponding two critical temperatures, \textit{for the case of IXs in a parabolic 2D trapping potential}, are
\begin{subequations} \label{eq:Tc}
\begin{align}
	T_{c1} &= \varepsilon_{bd}\sqrt{G/(A\gamma_{th})}, \\
    T_{c2} &= \varepsilon_{bd}\sqrt{(G-\gamma_{nr}N_{c2})/(A\gamma_{th})}.
\end{align}
\end{subequations}\\

Another peculiar dynamical effect, indicating the transitions between the different regimes, is predicted if the IX fluid is initialized at $t=0$ with population $\bar{N}_0>\bar{N}_{c2}$, and then freely decays. The solution of Eq.~\ref{eq:general_rate} for the time evolution of $\bar{N}$ is given in detail by supplementary Eq. S5. The results are illustrated by the black line in Fig.~\ref{fig:dynamics}(c). It reveals an initial fast exponential decay with a rate $\gamma_{th}$ of the mixed condensate, which slows down abruptly to $\gamma_{nr}$ at a time, marked as $t_1$, when the condensate becomes dark and then increases back again to $\gamma_{th}$ at a time $t_2$, when the condensate is fully depleted leaving only a thermal cloud. As a consequence, the observed emission intensity, proportional to $2N_B$, also displays a peculiar behavior, which is given by supplementary Eq. S9 and is also illustrated by the red dashed line in Fig.~\ref{fig:dynamics}(c). The surprising result here is that while the IX fluid is in the dark condensate phase, the photoluminescence is expected to be constant. This is understood since while in a dark condensate state, radiative recombination only occurs for particles occupying the thermal cloud, and the number of these is constant (see Eq.~\ref{eq:BEC_dist}) as long as the condensate exists. \vspace*{0.1cm}

\textbf{Discussion:} Strong non-monotonic growth of the BEC population with pump power or temperature, and the peculiar decay dynamics of the cloud are unique to a system with the particle number being a dynamical variable, rather than a conserved quantity as in the case of atomic BECs. It is due to a redistribution of particles between two states with very different decay times.   
Remarkably, the peculiar dependence of the total number of particles $\bar{N}$ on both $G$ and $T$ illustrated in Fig.~\ref{fig:dynamics}, is very similar to our recent experimental observations of a trapped IX fluid Ref.~\citenum{cohen_dark_2016,mazuz-harpaz_condensation_2017}. In that work we estimated $n_{c2}\simeq 3\text{-}8\times 10^{10}$cm$^{-2}$, which is very similar to the density of the dark-bright liquid transition estimated by the present model of a correlated liquid in Fig.~\ref{fig:nc2}. 
Furthermore, the fact that $n_{c2}$ depends only weakly on $\varepsilon_{bd}$ in the liquid might explain the robustness of the phase diagram to variations of the applied voltages observed in several experiments \cite{stern_exciton_2014,cohen_dark_2016,misra_experimental_2018}, in spite of the fact that such variations should in general have a significant effect on other parameters such as $\varepsilon_{bd}$ \cite{chen_effect_1987}.
\vspace*{0.1cm}

\textbf{Acknowledgements:} We greatly appreciate discussions with Boris Laikhtman. 
We acknowledge the support by the Israel Science Foundation, Grant No. 1287/15.
and Grant No. 836/17 and by the US - Israel Binational Science Foundation, grant No. 2016112.

\bibliography{My_Library}

\begin{thebibliography}{30}%
\makeatletter
\providecommand \@ifxundefined [1]{%
 \@ifx{#1\undefined}
}%
\providecommand \@ifnum [1]{%
 \ifnum #1\expandafter \@firstoftwo
 \else \expandafter \@secondoftwo
 \fi
}%
\providecommand \@ifx [1]{%
 \ifx #1\expandafter \@firstoftwo
 \else \expandafter \@secondoftwo
 \fi
}%
\providecommand \natexlab [1]{#1}%
\providecommand \enquote  [1]{``#1''}%
\providecommand \bibnamefont  [1]{#1}%
\providecommand \bibfnamefont [1]{#1}%
\providecommand \citenamefont [1]{#1}%
\providecommand \href@noop [0]{\@secondoftwo}%
\providecommand \href [0]{\begingroup \@sanitize@url \@href}%
\providecommand \@href[1]{\@@startlink{#1}\@@href}%
\providecommand \@@href[1]{\endgroup#1\@@endlink}%
\providecommand \@sanitize@url [0]{\catcode `\\12\catcode `\$12\catcode
  `\&12\catcode `\#12\catcode `\^12\catcode `\_12\catcode `\%12\relax}%
\providecommand \@@startlink[1]{}%
\providecommand \@@endlink[0]{}%
\providecommand \url  [0]{\begingroup\@sanitize@url \@url }%
\providecommand \@url [1]{\endgroup\@href {#1}{\urlprefix }}%
\providecommand \urlprefix  [0]{URL }%
\providecommand \Eprint [0]{\href }%
\providecommand \doibase [0]{http://dx.doi.org/}%
\providecommand \selectlanguage [0]{\@gobble}%
\providecommand \bibinfo  [0]{\@secondoftwo}%
\providecommand \bibfield  [0]{\@secondoftwo}%
\providecommand \translation [1]{[#1]}%
\providecommand \BibitemOpen [0]{}%
\providecommand \bibitemStop [0]{}%
\providecommand \bibitemNoStop [0]{.\EOS\space}%
\providecommand \EOS [0]{\spacefactor3000\relax}%
\providecommand \BibitemShut  [1]{\csname bibitem#1\endcsname}%
\let\auto@bib@innerbib\@empty
\bibitem [{\citenamefont {Baranov}\ \emph {et~al.}(2012)\citenamefont
  {Baranov}, \citenamefont {Dalmonte}, \citenamefont {Pupillo},\ and\
  \citenamefont {Zoller}}]{baranov_condensed_2012}%
  \BibitemOpen
  \bibfield  {author} {\bibinfo {author} {\bibfnamefont {M.~A.}\ \bibnamefont
  {Baranov}}, \bibinfo {author} {\bibfnamefont {M.}~\bibnamefont {Dalmonte}},
  \bibinfo {author} {\bibfnamefont {G.}~\bibnamefont {Pupillo}}, \ and\
  \bibinfo {author} {\bibfnamefont {P.}~\bibnamefont {Zoller}},\ }\href
  {\doibase 10.1021/cr2003568} {\bibfield  {journal} {\bibinfo  {journal}
  {Chemical Reviews}\ }\textbf {\bibinfo {volume} {112}},\ \bibinfo {pages}
  {5012} (\bibinfo {year} {2012})}\BibitemShut {NoStop}%
\bibitem [{\citenamefont {Lahaye}\ \emph {et~al.}(2009)\citenamefont {Lahaye},
  \citenamefont {Menotti}, \citenamefont {Santos}, \citenamefont {Lewenstein},\
  and\ \citenamefont {Pfau}}]{lahaye_physics_2009}%
  \BibitemOpen
  \bibfield  {author} {\bibinfo {author} {\bibfnamefont {T.}~\bibnamefont
  {Lahaye}}, \bibinfo {author} {\bibfnamefont {C.}~\bibnamefont {Menotti}},
  \bibinfo {author} {\bibfnamefont {L.}~\bibnamefont {Santos}}, \bibinfo
  {author} {\bibfnamefont {M.}~\bibnamefont {Lewenstein}}, \ and\ \bibinfo
  {author} {\bibfnamefont {T.}~\bibnamefont {Pfau}},\ }\href {\doibase
  10.1088/0034-4885/72/12/126401} {\bibfield  {journal} {\bibinfo  {journal}
  {Reports on Progress in Physics}\ }\textbf {\bibinfo {volume} {72}},\
  \bibinfo {pages} {126401} (\bibinfo {year} {2009})}\BibitemShut {NoStop}%
\bibitem [{\citenamefont {Kadau}\ \emph {et~al.}(2016)\citenamefont {Kadau},
  \citenamefont {Schmitt}, \citenamefont {Wenzel}, \citenamefont {Wink},
  \citenamefont {Maier}, \citenamefont {Ferrier-Barbut},\ and\ \citenamefont
  {Pfau}}]{kadau_observing_2016}%
  \BibitemOpen
  \bibfield  {author} {\bibinfo {author} {\bibfnamefont {H.}~\bibnamefont
  {Kadau}}, \bibinfo {author} {\bibfnamefont {M.}~\bibnamefont {Schmitt}},
  \bibinfo {author} {\bibfnamefont {M.}~\bibnamefont {Wenzel}}, \bibinfo
  {author} {\bibfnamefont {C.}~\bibnamefont {Wink}}, \bibinfo {author}
  {\bibfnamefont {T.}~\bibnamefont {Maier}}, \bibinfo {author} {\bibfnamefont
  {I.}~\bibnamefont {Ferrier-Barbut}}, \ and\ \bibinfo {author} {\bibfnamefont
  {T.}~\bibnamefont {Pfau}},\ }\href {\doibase 10.1038/nature16485} {\bibfield
  {journal} {\bibinfo  {journal} {Nature}\ }\textbf {\bibinfo {volume} {530}},\
  \bibinfo {pages} {194} (\bibinfo {year} {2016})}\BibitemShut {NoStop}%
\bibitem [{\citenamefont {Ferrier-Barbut}\ \emph {et~al.}(2016)\citenamefont
  {Ferrier-Barbut}, \citenamefont {Kadau}, \citenamefont {Schmitt},
  \citenamefont {Wenzel},\ and\ \citenamefont
  {Pfau}}]{ferrier-barbut_observation_2016}%
  \BibitemOpen
  \bibfield  {author} {\bibinfo {author} {\bibfnamefont {I.}~\bibnamefont
  {Ferrier-Barbut}}, \bibinfo {author} {\bibfnamefont {H.}~\bibnamefont
  {Kadau}}, \bibinfo {author} {\bibfnamefont {M.}~\bibnamefont {Schmitt}},
  \bibinfo {author} {\bibfnamefont {M.}~\bibnamefont {Wenzel}}, \ and\ \bibinfo
  {author} {\bibfnamefont {T.}~\bibnamefont {Pfau}},\ }\href {\doibase
  10.1103/PhysRevLett.116.215301} {\bibfield  {journal} {\bibinfo  {journal}
  {Physical Review Letters}\ }\textbf {\bibinfo {volume} {116}},\ \bibinfo
  {pages} {215301} (\bibinfo {year} {2016})}\BibitemShut {NoStop}%
\bibitem [{\citenamefont {Chomaz}\ \emph {et~al.}(2016)\citenamefont {Chomaz},
  \citenamefont {Baier}, \citenamefont {Petter}, \citenamefont {Mark},
  \citenamefont {W{\"a}chtler}, \citenamefont {Santos},\ and\ \citenamefont
  {Ferlaino}}]{chomaz_quantum-fluctuation-driven_2016}%
  \BibitemOpen
  \bibfield  {author} {\bibinfo {author} {\bibfnamefont {L.}~\bibnamefont
  {Chomaz}}, \bibinfo {author} {\bibfnamefont {S.}~\bibnamefont {Baier}},
  \bibinfo {author} {\bibfnamefont {D.}~\bibnamefont {Petter}}, \bibinfo
  {author} {\bibfnamefont {M.~J.}\ \bibnamefont {Mark}}, \bibinfo {author}
  {\bibfnamefont {F.}~\bibnamefont {W{\"a}chtler}}, \bibinfo {author}
  {\bibfnamefont {L.}~\bibnamefont {Santos}}, \ and\ \bibinfo {author}
  {\bibfnamefont {F.}~\bibnamefont {Ferlaino}},\ }\href {\doibase
  10.1103/PhysRevX.6.041039} {\bibfield  {journal} {\bibinfo  {journal}
  {Physical Review X}\ }\textbf {\bibinfo {volume} {6}},\ \bibinfo {pages}
  {041039} (\bibinfo {year} {2016})}\BibitemShut {NoStop}%
\bibitem [{\citenamefont {High}\ \emph
  {et~al.}(2012{\natexlab{a}})\citenamefont {High}, \citenamefont {Leonard},
  \citenamefont {Remeika}, \citenamefont {Butov}, \citenamefont {Hanson},\ and\
  \citenamefont {Gossard}}]{high_condensation_2012}%
  \BibitemOpen
  \bibfield  {author} {\bibinfo {author} {\bibfnamefont {A.~A.}\ \bibnamefont
  {High}}, \bibinfo {author} {\bibfnamefont {J.~R.}\ \bibnamefont {Leonard}},
  \bibinfo {author} {\bibfnamefont {M.}~\bibnamefont {Remeika}}, \bibinfo
  {author} {\bibfnamefont {L.~V.}\ \bibnamefont {Butov}}, \bibinfo {author}
  {\bibfnamefont {M.}~\bibnamefont {Hanson}}, \ and\ \bibinfo {author}
  {\bibfnamefont {A.~C.}\ \bibnamefont {Gossard}},\ }\href {\doibase
  10.1021/nl300983n} {\bibfield  {journal} {\bibinfo  {journal} {Nano Letters}\
  }\textbf {\bibinfo {volume} {12}},\ \bibinfo {pages} {2605} (\bibinfo {year}
  {2012}{\natexlab{a}})}\BibitemShut {NoStop}%
\bibitem [{\citenamefont {High}\ \emph
  {et~al.}(2012{\natexlab{b}})\citenamefont {High}, \citenamefont {Leonard},
  \citenamefont {Remeika}, \citenamefont {Butov}, \citenamefont {Hanson},\ and\
  \citenamefont {Gossard}}]{high_spontaneous_2012}%
  \BibitemOpen
  \bibfield  {author} {\bibinfo {author} {\bibfnamefont {A.~A.}\ \bibnamefont
  {High}}, \bibinfo {author} {\bibfnamefont {J.~R.}\ \bibnamefont {Leonard}},
  \bibinfo {author} {\bibfnamefont {M.}~\bibnamefont {Remeika}}, \bibinfo
  {author} {\bibfnamefont {L.~V.}\ \bibnamefont {Butov}}, \bibinfo {author}
  {\bibfnamefont {M.}~\bibnamefont {Hanson}}, \ and\ \bibinfo {author}
  {\bibfnamefont {A.~C.}\ \bibnamefont {Gossard}}\ }(\bibinfo  {publisher}
  {OSA},\ \bibinfo {year} {2012})\ p.\ \bibinfo {pages} {QTu3D.3}\BibitemShut
  {NoStop}%
\bibitem [{\citenamefont {Laikhtman}\ and\ \citenamefont
  {Rapaport}(2009{\natexlab{a}})}]{laikhtman_exciton_2009}%
  \BibitemOpen
  \bibfield  {author} {\bibinfo {author} {\bibfnamefont {B.}~\bibnamefont
  {Laikhtman}}\ and\ \bibinfo {author} {\bibfnamefont {R.}~\bibnamefont
  {Rapaport}},\ }\href {\doibase 10.1103/PhysRevB.80.195313} {\bibfield
  {journal} {\bibinfo  {journal} {Physical Review B}\ }\textbf {\bibinfo
  {volume} {80}},\ \bibinfo {pages} {195313} (\bibinfo {year}
  {2009}{\natexlab{a}})}\BibitemShut {NoStop}%
\bibitem [{\citenamefont {Laikhtman}\ and\ \citenamefont
  {Rapaport}(2009{\natexlab{b}})}]{laikhtman_correlations_2009}%
  \BibitemOpen
  \bibfield  {author} {\bibinfo {author} {\bibfnamefont {B.}~\bibnamefont
  {Laikhtman}}\ and\ \bibinfo {author} {\bibfnamefont {R.}~\bibnamefont
  {Rapaport}},\ }\href {\doibase 10.1209/0295-5075/87/27010} {\bibfield
  {journal} {\bibinfo  {journal} {EPL (Europhysics Letters)}\ }\textbf
  {\bibinfo {volume} {87}},\ \bibinfo {pages} {27010} (\bibinfo {year}
  {2009}{\natexlab{b}})}\BibitemShut {NoStop}%
\bibitem [{\citenamefont {Shilo}\ \emph {et~al.}(2013)\citenamefont {Shilo},
  \citenamefont {Cohen}, \citenamefont {Laikhtman}, \citenamefont {West},
  \citenamefont {Pfeiffer},\ and\ \citenamefont
  {Rapaport}}]{shilo_particle_2013}%
  \BibitemOpen
  \bibfield  {author} {\bibinfo {author} {\bibfnamefont {Y.}~\bibnamefont
  {Shilo}}, \bibinfo {author} {\bibfnamefont {K.}~\bibnamefont {Cohen}},
  \bibinfo {author} {\bibfnamefont {B.}~\bibnamefont {Laikhtman}}, \bibinfo
  {author} {\bibfnamefont {K.}~\bibnamefont {West}}, \bibinfo {author}
  {\bibfnamefont {L.}~\bibnamefont {Pfeiffer}}, \ and\ \bibinfo {author}
  {\bibfnamefont {R.}~\bibnamefont {Rapaport}},\ }\href {\doibase
  10.1038/ncomms3335} {\bibfield  {journal} {\bibinfo  {journal} {Nature
  Communications}\ }\textbf {\bibinfo {volume} {4}},\ \bibinfo {pages} {2335}
  (\bibinfo {year} {2013})}\BibitemShut {NoStop}%
\bibitem [{\citenamefont {Stern}\ \emph {et~al.}(2014)\citenamefont {Stern},
  \citenamefont {Umansky},\ and\ \citenamefont
  {Bar-Joseph}}]{stern_exciton_2014}%
  \BibitemOpen
  \bibfield  {author} {\bibinfo {author} {\bibfnamefont {M.}~\bibnamefont
  {Stern}}, \bibinfo {author} {\bibfnamefont {V.}~\bibnamefont {Umansky}}, \
  and\ \bibinfo {author} {\bibfnamefont {I.}~\bibnamefont {Bar-Joseph}},\
  }\href {\doibase 10.1126/science.1243409} {\bibfield  {journal} {\bibinfo
  {journal} {Science}\ }\textbf {\bibinfo {volume} {343}},\ \bibinfo {pages}
  {55} (\bibinfo {year} {2014})}\BibitemShut {NoStop}%
\bibitem [{\citenamefont {Cohen}\ \emph {et~al.}(2016)\citenamefont {Cohen},
  \citenamefont {Shilo}, \citenamefont {West}, \citenamefont {Pfeiffer},\ and\
  \citenamefont {Rapaport}}]{cohen_dark_2016}%
  \BibitemOpen
  \bibfield  {author} {\bibinfo {author} {\bibfnamefont {K.}~\bibnamefont
  {Cohen}}, \bibinfo {author} {\bibfnamefont {Y.}~\bibnamefont {Shilo}},
  \bibinfo {author} {\bibfnamefont {K.}~\bibnamefont {West}}, \bibinfo {author}
  {\bibfnamefont {L.}~\bibnamefont {Pfeiffer}}, \ and\ \bibinfo {author}
  {\bibfnamefont {R.}~\bibnamefont {Rapaport}},\ }\href {\doibase
  10.1021/acs.nanolett.6b01061} {\bibfield  {journal} {\bibinfo  {journal}
  {Nano Letters}\ }\textbf {\bibinfo {volume} {16}},\ \bibinfo {pages} {3726}
  (\bibinfo {year} {2016})}\BibitemShut {NoStop}%
\bibitem [{\citenamefont {Mazuz-Harpaz}\ \emph {et~al.}(2017)\citenamefont
  {Mazuz-Harpaz}, \citenamefont {Cohen},\ and\ \citenamefont
  {Rapaport}}]{mazuz-harpaz_condensation_2017}%
  \BibitemOpen
  \bibfield  {author} {\bibinfo {author} {\bibfnamefont {Y.}~\bibnamefont
  {Mazuz-Harpaz}}, \bibinfo {author} {\bibfnamefont {K.}~\bibnamefont {Cohen}},
  \ and\ \bibinfo {author} {\bibfnamefont {R.}~\bibnamefont {Rapaport}},\
  }\href {\doibase 10.1016/j.spmi.2017.01.022} {\bibfield  {journal} {\bibinfo
  {journal} {Superlattices and Microstructures}\ }\textbf {\bibinfo {volume}
  {108}},\ \bibinfo {pages} {88} (\bibinfo {year} {2017})}\BibitemShut
  {NoStop}%
\bibitem [{\citenamefont {Misra}\ \emph {et~al.}(2018)\citenamefont {Misra},
  \citenamefont {Stern}, \citenamefont {Joshua}, \citenamefont {Umansky},\ and\
  \citenamefont {Bar-Joseph}}]{misra_experimental_2018}%
  \BibitemOpen
  \bibfield  {author} {\bibinfo {author} {\bibfnamefont {S.}~\bibnamefont
  {Misra}}, \bibinfo {author} {\bibfnamefont {M.}~\bibnamefont {Stern}},
  \bibinfo {author} {\bibfnamefont {A.}~\bibnamefont {Joshua}}, \bibinfo
  {author} {\bibfnamefont {V.}~\bibnamefont {Umansky}}, \ and\ \bibinfo
  {author} {\bibfnamefont {I.}~\bibnamefont {Bar-Joseph}},\ }\href {\doibase
  10.1103/PhysRevLett.120.047402} {\bibfield  {journal} {\bibinfo  {journal}
  {Physical Review Letters}\ }\textbf {\bibinfo {volume} {120}},\ \bibinfo
  {pages} {047402} (\bibinfo {year} {2018})}\BibitemShut {NoStop}%
\bibitem [{\citenamefont {Anankine}\ \emph {et~al.}(2017)\citenamefont
  {Anankine}, \citenamefont {Beian}, \citenamefont {Dang}, \citenamefont
  {Alloing}, \citenamefont {Cambril}, \citenamefont {Merghem}, \citenamefont
  {Carbonell}, \citenamefont {Lema{\^i}tre},\ and\ \citenamefont
  {Dubin}}]{anankine_quantized_2017}%
  \BibitemOpen
  \bibfield  {author} {\bibinfo {author} {\bibfnamefont {R.}~\bibnamefont
  {Anankine}}, \bibinfo {author} {\bibfnamefont {M.}~\bibnamefont {Beian}},
  \bibinfo {author} {\bibfnamefont {S.}~\bibnamefont {Dang}}, \bibinfo {author}
  {\bibfnamefont {M.}~\bibnamefont {Alloing}}, \bibinfo {author} {\bibfnamefont
  {E.}~\bibnamefont {Cambril}}, \bibinfo {author} {\bibfnamefont
  {K.}~\bibnamefont {Merghem}}, \bibinfo {author} {\bibfnamefont {C.~G.}\
  \bibnamefont {Carbonell}}, \bibinfo {author} {\bibfnamefont {A.}~\bibnamefont
  {Lema{\^i}tre}}, \ and\ \bibinfo {author} {\bibfnamefont {F.}~\bibnamefont
  {Dubin}},\ }\href {\doibase 10.1103/PhysRevLett.118.127402} {\bibfield
  {journal} {\bibinfo  {journal} {Physical Review Letters}\ }\textbf {\bibinfo
  {volume} {118}},\ \bibinfo {pages} {127402} (\bibinfo {year}
  {2017})}\BibitemShut {NoStop}%
\bibitem [{\citenamefont {High}\ \emph {et~al.}(2013)\citenamefont {High},
  \citenamefont {Hammack}, \citenamefont {Leonard}, \citenamefont {Yang},
  \citenamefont {Butov}, \citenamefont {Ostatnicky}, \citenamefont
  {Vladimirova}, \citenamefont {Kavokin}, \citenamefont {Liew}, \citenamefont
  {Campman},\ and\ \citenamefont {Gossard}}]{high_spin_2013}%
  \BibitemOpen
  \bibfield  {author} {\bibinfo {author} {\bibfnamefont {A.~A.}\ \bibnamefont
  {High}}, \bibinfo {author} {\bibfnamefont {A.~T.}\ \bibnamefont {Hammack}},
  \bibinfo {author} {\bibfnamefont {J.~R.}\ \bibnamefont {Leonard}}, \bibinfo
  {author} {\bibfnamefont {S.}~\bibnamefont {Yang}}, \bibinfo {author}
  {\bibfnamefont {L.~V.}\ \bibnamefont {Butov}}, \bibinfo {author}
  {\bibfnamefont {T.}~\bibnamefont {Ostatnicky}}, \bibinfo {author}
  {\bibfnamefont {M.}~\bibnamefont {Vladimirova}}, \bibinfo {author}
  {\bibfnamefont {A.~V.}\ \bibnamefont {Kavokin}}, \bibinfo {author}
  {\bibfnamefont {T.~C.~H.}\ \bibnamefont {Liew}}, \bibinfo {author}
  {\bibfnamefont {K.~L.}\ \bibnamefont {Campman}}, \ and\ \bibinfo {author}
  {\bibfnamefont {A.~C.}\ \bibnamefont {Gossard}},\ }\href {\doibase
  10.1103/PhysRevLett.110.246403} {\bibfield  {journal} {\bibinfo  {journal}
  {Physical Review Letters}\ }\textbf {\bibinfo {volume} {110}},\ \bibinfo
  {pages} {246403} (\bibinfo {year} {2013})}\BibitemShut {NoStop}%
\bibitem [{\citenamefont {Beian}\ \emph {et~al.}(2017)\citenamefont {Beian},
  \citenamefont {Alloing}, \citenamefont {Anankine}, \citenamefont {Cambril},
  \citenamefont {Carbonell}, \citenamefont {{Aristide Lema{\^i}tre}},\ and\
  \citenamefont {Dubin}}]{beian_spectroscopic_2017}%
  \BibitemOpen
  \bibfield  {author} {\bibinfo {author} {\bibfnamefont {M.}~\bibnamefont
  {Beian}}, \bibinfo {author} {\bibfnamefont {M.}~\bibnamefont {Alloing}},
  \bibinfo {author} {\bibfnamefont {R.}~\bibnamefont {Anankine}}, \bibinfo
  {author} {\bibfnamefont {E.}~\bibnamefont {Cambril}}, \bibinfo {author}
  {\bibfnamefont {C.~G.}\ \bibnamefont {Carbonell}}, \bibinfo {author}
  {\bibnamefont {{Aristide Lema{\^i}tre}}}, \ and\ \bibinfo {author}
  {\bibfnamefont {F.}~\bibnamefont {Dubin}},\ }\href {\doibase
  10.1209/0295-5075/119/37004} {\bibfield  {journal} {\bibinfo  {journal} {EPL
  (Europhysics Letters)}\ }\textbf {\bibinfo {volume} {119}},\ \bibinfo {pages}
  {37004} (\bibinfo {year} {2017})}\BibitemShut {NoStop}%
\bibitem [{\citenamefont {Combescot}\ \emph {et~al.}(2007)\citenamefont
  {Combescot}, \citenamefont {Betbeder-Matibet},\ and\ \citenamefont
  {Combescot}}]{combescot_bose-einstein_2007}%
  \BibitemOpen
  \bibfield  {author} {\bibinfo {author} {\bibfnamefont {M.}~\bibnamefont
  {Combescot}}, \bibinfo {author} {\bibfnamefont {O.}~\bibnamefont
  {Betbeder-Matibet}}, \ and\ \bibinfo {author} {\bibfnamefont
  {R.}~\bibnamefont {Combescot}},\ }\href {\doibase
  10.1103/PhysRevLett.99.176403} {\bibfield  {journal} {\bibinfo  {journal}
  {Physical Review Letters}\ }\textbf {\bibinfo {volume} {99}},\ \bibinfo
  {pages} {176403} (\bibinfo {year} {2007})}\BibitemShut {NoStop}%
\bibitem [{\citenamefont {Combescot}\ and\ \citenamefont
  {Combescot}(2012)}]{combescot_``gray_2012}%
  \BibitemOpen
  \bibfield  {author} {\bibinfo {author} {\bibfnamefont {R.}~\bibnamefont
  {Combescot}}\ and\ \bibinfo {author} {\bibfnamefont {M.}~\bibnamefont
  {Combescot}},\ }\href {\doibase 10.1103/PhysRevLett.109.026401} {\bibfield
  {journal} {\bibinfo  {journal} {Physical Review Letters}\ }\textbf {\bibinfo
  {volume} {109}},\ \bibinfo {pages} {026401} (\bibinfo {year}
  {2012})}\BibitemShut {NoStop}%
\bibitem [{Note1()}]{Note1}%
  \BibitemOpen
  \bibinfo {note} {The two independent experiments which reported the
  observation of that robust dark liquid were performed on IXs with different
  and relatively large dipole lengths: $16$nm \cite {cohen_dark_2016} and
  $18$nm \cite {stern_exciton_2014,misra_experimental_2018}}\BibitemShut
  {NoStop}%
\bibitem [{\citenamefont {Pitaevskii}\ and\ \citenamefont
  {Stringari}(2003)}]{pitaevskii_boseeinstein_2003}%
  \BibitemOpen
  \bibfield  {author} {\bibinfo {author} {\bibfnamefont {L.}~\bibnamefont
  {Pitaevskii}}\ and\ \bibinfo {author} {\bibfnamefont {S.}~\bibnamefont
  {Stringari}},\ }\href@noop {} {\emph {\bibinfo {title}
  {Bose{\textendash}{Einstein} condensation}}}\ (\bibinfo  {publisher} {Oxford
  University Press (Oxford, 2003)},\ \bibinfo {year} {2003})\BibitemShut
  {NoStop}%
\bibitem [{Note2()}]{Note2}%
  \BibitemOpen
  \bibinfo {note} {In a two-dimensional parabolic trap and for a
  non-interacting system, $B(T,\varepsilon _{bd})= A(T/\varepsilon _{bd} )^2$,
  where $A$ is a constant, and we note that in the case of interactions, the
  transition becomes of a BKT type into a superfluid state, with a different
  temperature dependence. In this work however, we do not discuss the details
  of the phase transition into the condensed phase, but rather assume such a
  phase has been reached.}\BibitemShut {Stop}%
\bibitem [{\citenamefont {Zimmermann}\ and\ \citenamefont
  {Schindler}(2007)}]{zimmermann_excitonexciton_2007}%
  \BibitemOpen
  \bibfield  {author} {\bibinfo {author} {\bibfnamefont {R.}~\bibnamefont
  {Zimmermann}}\ and\ \bibinfo {author} {\bibfnamefont {C.}~\bibnamefont
  {Schindler}},\ }\href {\doibase 10.1016/j.ssc.2007.07.044} {\bibfield
  {journal} {\bibinfo  {journal} {Solid State Communications}\ }\textbf
  {\bibinfo {volume} {144}},\ \bibinfo {pages} {395} (\bibinfo {year}
  {2007})}\BibitemShut {NoStop}%
\bibitem [{Note3()}]{Note3}%
  \BibitemOpen
  \bibinfo {note} {Defined this way, $\Phi _d$ is approximately normalized over
  the entire area of the system $L^2$, neglecting the effect of the scattering
  on the wave-function at distances exceeding $k^{-1}$}\BibitemShut {NoStop}%
\bibitem [{Note4()}]{Note4}%
  \BibitemOpen
  \bibinfo {note} {For IXs in a condensate the typical wavelength is of the
  order of $2\pi /k\sim L\gg a_X$. In addition, the dependence of $\Phi _d$ on
  $k$ is logarithmic and therefore weak. Therefore, for $2\pi /k\gg a_X$ \cite
  {laikhtman_exciton_2009}, we can safely choose $k^{-1}=10 a_X$ to calculate
  $\xi _d$ as a function of $d$. See supplementary note S1.}\BibitemShut
  {Stop}%
\bibitem [{\citenamefont {Pikus}\ and\ \citenamefont
  {Bir}(1971)}]{pikus_exchange_1971}%
  \BibitemOpen
  \bibfield  {author} {\bibinfo {author} {\bibfnamefont {G.}~\bibnamefont
  {Pikus}}\ and\ \bibinfo {author} {\bibfnamefont {G.}~\bibnamefont {Bir}},\
  }\href@noop {} {\bibfield  {journal} {\bibinfo  {journal} {Soviet Physics
  JETP}\ }\textbf {\bibinfo {volume} {33}},\ \bibinfo {pages} {108} (\bibinfo
  {year} {1971})}\BibitemShut {NoStop}%
\bibitem [{\citenamefont {Chen}\ \emph {et~al.}(1988)\citenamefont {Chen},
  \citenamefont {Gil}, \citenamefont {Lefebvre},\ and\ \citenamefont
  {Mathieu}}]{chen_exchange_1988}%
  \BibitemOpen
  \bibfield  {author} {\bibinfo {author} {\bibfnamefont {Y.}~\bibnamefont
  {Chen}}, \bibinfo {author} {\bibfnamefont {B.}~\bibnamefont {Gil}}, \bibinfo
  {author} {\bibfnamefont {P.}~\bibnamefont {Lefebvre}}, \ and\ \bibinfo
  {author} {\bibfnamefont {H.}~\bibnamefont {Mathieu}},\ }\href {\doibase
  10.1103/PhysRevB.37.6429} {\bibfield  {journal} {\bibinfo  {journal}
  {Physical Review B}\ }\textbf {\bibinfo {volume} {37}},\ \bibinfo {pages}
  {6429} (\bibinfo {year} {1988})}\BibitemShut {NoStop}%
\bibitem [{Note5()}]{Note5}%
  \BibitemOpen
  \bibinfo {note} {We verified that the contribution from the exciton's
  neighbors other than nearest is negligible, since the exchange integral
  decays exponentially with distance, $r$ on the length scale of the IX
  radius.}\BibitemShut {Stop}%
\bibitem [{Note6()}]{Note6}%
  \BibitemOpen
  \bibinfo {note} {We note that some dependence on $d$ should appear if the
  harmonic approximation for $\Phi _d$ from Ref.~\protect \citenum
  {laikhtman_exciton_2009} instead of the Delta function of Eq.~\ref
  {eq:Phi_fixed}.}\BibitemShut {Stop}%
\bibitem [{\citenamefont {Chen}\ \emph {et~al.}(1987)\citenamefont {Chen},
  \citenamefont {Koteles}, \citenamefont {Elman},\ and\ \citenamefont
  {Armiento}}]{chen_effect_1987}%
  \BibitemOpen
  \bibfield  {author} {\bibinfo {author} {\bibfnamefont {Y.~J.}\ \bibnamefont
  {Chen}}, \bibinfo {author} {\bibfnamefont {E.~S.}\ \bibnamefont {Koteles}},
  \bibinfo {author} {\bibfnamefont {B.~S.}\ \bibnamefont {Elman}}, \ and\
  \bibinfo {author} {\bibfnamefont {C.~A.}\ \bibnamefont {Armiento}},\ }\href
  {\doibase 10.1103/PhysRevB.36.4562} {\bibfield  {journal} {\bibinfo
  {journal} {Physical Review B}\ }\textbf {\bibinfo {volume} {36}},\ \bibinfo
  {pages} {4562} (\bibinfo {year} {1987})}\BibitemShut {NoStop}%
\end{thebibliography}%

\end{document}